\documentstyle[preprint,prc,aps,floats,epsf]{revtex}
\def\thalf{{\textstyle{\frac{1}{2}}}}
\def\thrhalf{{\textstyle{\frac{3}{2}}}}
\def\toneth{{\textstyle{\frac{1}{3}}}}

\begin{document}


\preprint{NUC-MINN-96/4-T}

\title{Redundance of $\Delta$-isobar Parameters in Effective Field Theories}

\author{Hua-Bin Tang and Paul J. Ellis}
\address{School of Physics and Astronomy,
     University of Minnesota, Minneapolis, MN\ \ 55455.}

\date{March 14, 1996}
\maketitle

\begin{abstract}
It is shown that the off-shell parameters
in the interaction Lagrangian of pions, nucleons, and 
$\Delta$-isobars are redundant in the framework of effective
field theories. Our results also suggest the
necessity of including the $\Delta$ as an explicit dynamical degree of freedom.
\end{abstract}

\vspace{20pt}
%


The spin-$\thrhalf$ isospin-$\thrhalf$ $\Delta$ isobar 
plays an  important role in fitting pion-nucleon and nucleon-nucleon 
scattering data. However,
a satisfactory covariant treatment of spin-$\thrhalf$ fields has been hampered
\cite{HOHLER83}
by the existence of two parameters commonly called $A$ and $Z$.
Nath et al.\cite{NATH71} have argued that physical quantities are 
independent of $A$,
while they depend on the off-shell parameter $Z$ which occurs in the
$\pi N\Delta$ interaction. They further suggested that quantum theory 
requires $Z=\thalf$. Olsson and Osypowski\cite{OLSSON75} mentioned
another claim that both $Z=\pm\thalf$ are possible.
Various choices of the $Z$ parameter,
including fitting it to experimental data,
have been discussed in Ref.\cite{BEN89}. In a recent fit\cite{GOUDSMIT94} to
$\pi N$ scattering data, a value of $Z\sim-\thalf$ was favored.

Notice that the above models were not in the framework of
modern effective field theories (EFTs)\cite{EFTrefs},
of which chiral perturbation theory
(ChPT) \cite{W79GL84} is a successful prototype.
Nevertheless, the role of the off-shell
parameters in modern EFTs is also of interest.
Recently, Banerjee and Milana\cite{BAN_MIL95} have examined
$1/M$ corrections (where $M$ is a typical baryon mass)
in heavy-baryon chiral perturbation theory (ChPT)\cite{JENKINS91}.
They suggested, on the one hand, 
that one may be at an impasse without knowledge of the off-shell
parameters and, on the other hand, that the ``off-shell''
$\Delta$ effects could be absorbed as part of the higher-dimension 
operators using purely ``on-shell'' baryon fields.
In this Letter,
we show  that the off-shell parameter
in the $\pi N\Delta$ interaction, and similar parameters which arise for
the $\pi \Delta \Delta$ interaction,
can be subsumed in other terms in the effective Lagrangian.
We demonstate further that the $\Delta$ isobar should be treated 
as a dynamical degree of freedom in agreement with
the view taken in e.g.
Refs.~\cite{BAN_MIL95,JENKINS91b,BIRA},
but at variance with
that of Bernard et al.~\cite{MEISSNER}.

The free Lagrangian for the spin-$\thrhalf$ and isospin-$\thrhalf$ 
vector-spinor field
\begin{equation}
\Delta_\mu = \left( 
                    \begin{array}{l}
                       \Delta_\mu^{++}\\
                       \Delta_\mu^{+}\\
                       \Delta_\mu^{0}\\
                       \Delta_\mu^{-}
                    \end{array}
             \right)       \label{eq:Del}
\end{equation}
is given by\cite{MOLDAUER56}
\begin{equation}
  {\cal{L}}_0 = \overline{\Delta}_\mu \Lambda^{\mu\nu}_0\Delta_\nu \ ,
\end{equation}
where\footnote{We use the conventions of Bjorken and Drell\cite{BD65}.}
\begin{eqnarray}
  \Lambda^{\mu\nu}_0 &=& -(i\rlap/{\partial} - M_\Delta)g^{\mu\nu}
       - i A (\gamma^\mu \partial^\nu + \gamma^\nu\partial^\mu)
                       \nonumber \\
    & & 
    -\thalf i (3A^2+2A+1)\gamma^\mu \rlap/{\partial} \gamma^\nu
       -(3A^2+3A+1) M_\Delta \gamma^\mu \gamma^\nu \ ,
                         \label{eq:lambda}
\end{eqnarray}
with $A\ (\neq-\thalf)$ a parameter.
Note the difference in an overall minus sign from some recent 
work\cite{BEN89,GRIEGEL91}, as already noticed in 
Ref.~\cite{BAN_MIL95}. 
This ensures that the physical spatial components of
the $\Delta$ field behave correctly like a Dirac field and the resulting
Hamiltonian is positive definite. 
The Lagrangian is invariant under the so-called point
transformation\cite{NATH71},
\begin{eqnarray}
  \Delta_\mu &\rightarrow& \Delta'_\mu
            =\Delta_\mu + a \gamma_\mu\gamma_\nu \Delta^\nu 
                       \ ,\nonumber\\
  A &\rightarrow& A' = {A-2a \over 1+4a} \ ,\label{eq:fred}
\end{eqnarray}
where $a$ is arbitrary except $a\neq -1/4$.
Note that the above transformation is {\it not} a symmetry of
the Lagrangian since it changes the parameter $A$. 
Thus, it is more precise to call it
a field redefinition which does not change physical 
quantities and, as a result, the parameter $A$ 
has no physical consequences.

The low-energy Lagrangian for the $\pi N \Delta$ system 
can be written as an expansion
in derivatives of the pion field. Chiral symmetry ($SU(2)\otimes SU(2)$
realized in the 
non-linear form), Lorentz invariance and
parity dictate the form of the Lagrangian which can be written as
the sum of two parts, one (${\cal L}_{\pi N}$)
involves only pions and nucleons and
the other (${\cal L}_\Delta $) also contains $\Delta$ isobars. 
That is
\begin{equation}
  {\cal L} = {\cal L}_{\pi N} + {\cal L}_\Delta \ . \label{eq:L}
\end{equation}

Up to two derivatives on the pion field, we have
\begin{eqnarray}
  {\cal L}_{\pi N} &=& \overline{N} ( i\rlap/{\mkern-2mu\cal D}+g_{\rm A} 
             \gamma^\mu \gamma_5 a_\mu - M ) N 
+{f_\pi^2 \over 4} {\rm tr}\, (\partial_\mu U^\dagger \partial^\mu U)
                  \nonumber  \\
      & & +{1\over 4}m_\pi^2f_\pi^2 {\rm tr}\,(U + U^\dagger -2)
         + {1\over M} \overline{N} \Big(
         \beta_\pi {\rm tr}\, (\partial_\mu U^\dagger \partial^\mu U)
          -\kappa_\pi v_{\mu\nu}\sigma^{\mu\nu} \Big ) N
                     \nonumber  \\
   & & +\frac{\kappa_1}{2M^2}\left[i\overline{N}\gamma_{\mu}{\cal D}_{\nu} N\,
    {\rm tr}\,(a^{\mu}a^{\nu})+h.c.\right] 
    +{\kappa_2\over  M} m_\pi^2 \overline{N} N\, 
            {\rm tr}\,(U + U^\dagger -2) 
                \ ,     \label{eq:LpiN}
\end{eqnarray}
where the nucleon covariant derivative is
${\cal D}_\mu N = \partial_\mu N + i v_\mu N$,
the trace ``tr'' is over $SU(2)$ isospin indices,
$f_\pi \approx 93\,$MeV is the pion decay constant,
$g_{\rm A} \approx 1.26$ is the axial coupling, 
and $\beta_\pi$, $\kappa_\pi$, $\kappa_1$ and $\kappa_2$ are additional
higher-order couplings. 
Note that we have organized the terms in accordance with
Georgi's naive dimensional analysis\cite{GEORGI}.
In addition to the chiral invariant terms, we have introduced 
symmetry-breaking terms proportional to the square of the pion mass 
$m_\pi^2$.
The axial and vector fields are defined in terms of the pion field by
\begin{eqnarray}
a_{\mu}     & \equiv & -{i \over 2}(\xi^{\dagger} \partial_{\mu} \xi -
    \xi \partial_{\mu}\xi^{\dagger} )
          = a_{\mu}^\dagger
          = \thalf\bbox{a}_\mu\bbox{\cdot\tau}
          =\frac{1}{f_{\pi}}\partial_{\mu}\pi+\cdots
                \ ,\label{eq:adef}\\[4pt]
v_{\mu}     & \equiv &
           -{i \over 2}(\xi^{\dagger} \partial_{\mu} \xi +
    \xi \partial_{\mu}\xi^{\dagger} )
         = v_{\mu} ^\dagger
         = \thalf\bbox{v}_\mu\bbox{\cdot\tau}
         = -\frac{i}{2f^2_{\pi}}[\pi,\partial_{\mu}\pi]+\cdots
            \ , \label{eq:vdef}\\[4pt]
v_{\mu\nu} &=& -i[a_{\mu},a_{\nu}]=\partial_{\mu} v_{\nu} -\partial_{\nu}
      v_{\mu} + i [v_{\mu}, v_{\nu}] \ ,
\end{eqnarray}
with $\pi \equiv \thalf\bbox{\pi\cdot \tau}$ and
$U(x) = \xi \xi =\exp (2i\pi(x)/ f_{\pi})$.

The part of the chiral Lagrangian involving the $\Delta$ is
\begin{equation}
  {\cal L}_\Delta    =  
  \overline{\Delta}_\mu^a
           \Lambda^{\mu\nu}_{ab} \Delta_\nu^b 
       + h_{\rm A} \Big ( \overline{\bbox{\Delta}}_\mu 
         \bbox{\cdot a}_\nu \Theta^{\mu\nu} N
       + \overline{N}
           \Theta^{\mu\nu} \bbox{a}_\mu\bbox{\cdot \Delta}_\nu \Big)
         + \tilde{h}_{\rm A} 
             \overline{\Delta}_\mu^{\, a}
                  \Gamma^{\mu\nu\sigma} \gamma_5 a_\sigma               
                    \Delta_\nu^a 
              \ ,              \label{eq:lagDel}
\end{equation}
where  $h_{\rm A}$
and $\tilde{h}_{\rm A}$
are the $\pi N \Delta$ and $\pi \Delta \Delta$ couplings.
We have found it convenient to define the $\Delta$ in terms of three 
two-component spinors in isospin space (see the Appendix) by setting 
$\bbox{\Delta}_\mu = \bbox{T} \Delta_\mu$
using the standard $2\times 4$
isospin $\thrhalf$ to $\thalf$  transition matrix:
\begin{equation}
\langle {\case1/2}\, t |
   \bbox{T} |{\case3/2}\, t_\Delta \rangle
  \equiv \sum_\lambda \langle 1\, \lambda \,{\case1/2} \,t |
    {\case3/2} \,t_\Delta\rangle \bbox {e}_\lambda
                     \ , \label{eq:T}
\end{equation}
with the isospin spherical unit vectors $\bbox{e}_0 = \bbox{e}_z$ and
$\bbox{e}_{\pm 1}=\mp (\bbox{e}_x \pm i \bbox{e}_y)/\sqrt{2}$. 
The kernel in the first term of (\ref{eq:lagDel}), 
$\Lambda^{\mu\nu}_{ab}$, is obtained from $\Lambda^{\mu\nu}_0\delta_{ab}$
of (\ref{eq:lambda}) 
by replacing the ordinary
derivative with the covariant derivative
\begin{equation}
{\cal D}_\mu \bbox{\Delta}_\nu 
            = \partial_\mu\bbox{\Delta}_\nu
                      + i  v_\mu  \bbox{\Delta}_\nu
                      -   \bbox{v}_\mu \times \bbox{\Delta}_\nu
                            \ . \label{eq:covder}
\end{equation}
We note in passing that
\begin{equation}
\bbox{T}^{\dagger}\bbox{\cdot}{\cal D}_\mu \bbox{\Delta}_\nu=
\partial_\mu{\Delta}_\nu
+i\bbox{v}_{\mu}\bbox{\cdot t}^{\left(\frac{3}{2}\right)}\Delta_{\nu}\;,
\end{equation}
where $\bbox{t}^{\left(\frac{3}{2}\right)}$ are the generators of the 
isospin-$\thrhalf$ algebra.
For the $\pi N\Delta$ term we employ the standard definition\cite{NATH71}
\begin{equation}
  \Theta_{\mu\nu} = g_{\mu\nu} +\thalf
        [(4 Z + 1) A + 2 Z]\gamma_\mu\gamma_\nu 
             \ , \label{thetadef}
\end{equation}
which defines the off-shell $Z$ parameter. For the  $\pi\Delta\Delta$ 
interaction we have defined
\begin{eqnarray}
     \Gamma^{\mu\nu\sigma} &\equiv&
             g^{\mu\nu} \gamma^\sigma
           +   [(4 Z_2 +1 ) A + 2 Z_2] 
           ( \gamma^\mu g^{\nu \sigma}
                  + \gamma^\nu g^{\mu \sigma} )
                    \nonumber  \\
      & &    - \Big[ \Big(2 Z_2 + \thalf\Big) A +  Z_2
                + \Big(4 Z_3 + \thalf\Big) (A^2+A) + Z_3
                       \Big]  \gamma^\mu
                \gamma^\sigma  \gamma^\nu \ , 
\end{eqnarray}
which involves two additional off-shell parameters $Z_2$ and $Z_3$.
Each of the three terms in ${\cal L}_{\Delta}$ is invariant under the field 
redefinition (\ref{eq:fred}), indeed the $Z$ parameters have been 
introduced precisely to preserve this invariance.
To make this Letter self-contained, the chiral invariance of the Lagrangian 
warrants a brief discussion; this is deferred to the Appendix.

In what follows we show that, besides the $A$ parameter, 
the dependence on the $Z$
parameters can also be absorbed in the other parameters in the
Lagrangian. To achieve this, we integrate out the $\Delta$ 
field and examine the resulting equivalent $\pi N$ Lagrangian.
This can be done by performing the 
standard Gaussian integration since
the Lagrangian
(\ref{eq:lagDel}) is quadratic in the $\Delta$ field to the order we
keep.  Because the off-shell vertices connected to on-shell
external lines vanish, it is sufficient to discuss the case 
without an external source for the $\Delta$.

First, in the standard functional integral, we may redefine the 
$\Delta$ field as in  (\ref{eq:fred}) with
$a=-(A+1)/2$ so that $A' = -1$.
Since the Jacobian is constant we see explicitly that the dependence
on the parameter $A$ is removed.
Now we may formally identify
\begin{equation}
\eta_\mu(x) = h_{\rm A} \Theta_{\mu\nu}
            \bbox{T}^\dagger\bbox{\cdot a}^\nu N 
\end{equation}
as a ``source" for the $\Delta$ field.
Performing the $\Delta$ functional integration
then yields an equivalent  Lagrangian
\begin{eqnarray}
  \int d^4 x {\cal L}_{\equiv} &=& 
                \int d^4\! x\, {\cal L}_{\pi N} 
              - i {\rm Tr} \ln (K \Lambda^{-1}_0)
                         \nonumber  \\
      & &
      -\int d^4\!x d^4 \! y\, \overline{\eta}^\mu(x)
                G_{\mu\nu}(x,y) \eta^\nu(y)
            \ , \label{eq:effL}
\end{eqnarray}
where, in the second term, the trace ``Tr'' is over spacetime, isospin, and 
vector-spinor indices
and it has been normalized to vanish when the pion field is set to zero.
The inverse of the free propagator is given by \cite{BEN89,GRIEGEL91,PILKUHN}
\begin{equation}
 (\Lambda^{-1}_0)_{\mu\nu}(x,y) = S_{\mu\nu}(x)\delta^4(x-y) 
              \ , \label{eq:pgt}
\end{equation}
with
\begin{equation}
  S_{\mu\nu}(x) = {1\over i\rlap/\partial - M_\Delta+i\epsilon}
       \Big [ -g_{\mu\nu} +\toneth\gamma_\mu\gamma_\nu
              + {i\over 3 M_\Delta}(\gamma_\mu\partial_\nu-
             \gamma_\nu\partial_\mu) -
             {2\over 3 M_\Delta^2}\partial_\mu\partial_\nu \Big]
                  \ ,\label{eq:freep}
\end{equation}
and $G_{\mu\nu}(x,y)$ is the inverse of the kernal
\begin{equation}
K^{\mu\nu}(x,y) \equiv  T_a^\dagger\Big(
                  \Lambda^{\mu\nu}_{ab} 
                   +  \tilde{h}_{\rm A}
                \Gamma^{\mu\nu\sigma}
                \gamma_5 a_\sigma \delta_{ab} \Big)T_b  \delta^4(x-y)   \ .
                         \label{eq:K}
\end{equation}

The trace term in (\ref{eq:effL}) corresponds to short-distance physics
since the mass of the $\Delta$ sets the relevant scale. This trace
can be expanded in derivatives of the pion field, resulting in 
local terms that can be absorbed into
the coefficients in ${\cal L}_{\pi N}$ as in 
Ref.\cite{FTS95}. 
Up to two  derivatives on the pion field, one can replace 
$G_{\mu\nu}(x,y) $ in the last term of (\ref{eq:effL}) with
the free propagator given in (\ref{eq:freep}).
This is the case since 
the terms involving the pion field
in the denominator of $G_{\mu\nu}(x,y) $
can be treated perturbatively as a result of the fact that 
a derivative on the pion field is assciated with  
a  factor of $1/(4\pi f_\pi)$ from the loop integral
\cite{MANGEORGI}.
The corresponding contribution to the
Lagrangian ${\cal L}_{\equiv}$ can then be written as
\begin{equation}
{\cal L}_1 (x)   =   
              - h_{\rm A}^2  \overline{N} 
             \bbox{T\cdot a}_\mu
           \Theta^{\mu\lambda}S_{\lambda\sigma}
           \Theta^{\sigma\nu}
            \bbox{T}^\dagger\bbox{\cdot a}_\nu N
              \ . \label{eq:L1}
\end{equation}
Noticing that
\begin{equation}
\gamma_\mu S^{\mu\nu} = S^{\nu\mu}\gamma_\mu 
       = {1\over 3 M_\Delta}
         \Big ( {2\over  M_\Delta} i \partial^\nu -\gamma^\nu \Big)
                     \ , \label{eq:red}
\end{equation}
one finds
\begin{eqnarray}
\Theta_{\mu\lambda}S^{\lambda\sigma}\Theta_{\sigma\nu}
           &=& S_{\mu\nu}    +{2\over 3M_\Delta^2}(Z+\case1/2)
             [M_\Delta\gamma_\mu\gamma_\nu-(\gamma_\mu i\partial_\nu
                     +\gamma_\nu i \partial_\mu)] 
                      \nonumber \\
       & & \ \ \ \ -  {2\over 3M_\Delta^2}(Z+\case1/2)^2
             \gamma_\mu(2 M_\Delta - i\rlap/\partial)\gamma_\nu
                           \ . \label{eq:multout}
\end{eqnarray}

Putting (\ref{eq:multout}) into 
(\ref{eq:L1}) and using the nucleon equation of motion to remove
derivatives on the nucleon field, we find
\begin{eqnarray}
{\cal L}_1(x)   & =&
              - h_{\rm A}^2  \overline{N}
             \bbox{T\cdot a}_\mu S^{\mu\nu}
            \bbox{T}^\dagger\bbox{\cdot a}_\nu N
         -{8\over 9} {h_{\rm A}^2 \over M_\Delta^2}
          (Z^2-\case1/4)\, \left[i\overline{N}\,
            \gamma_\mu  \partial_\nu N {\rm tr}\, (a^\mu a^\nu)+h.c.\right]
             \nonumber \\  
     & &
          + C [  \overline{N} N\,{\rm tr}\, (
            \partial_\mu U^\dagger \partial^\mu U)
           - 2\overline{N} v_{\mu\nu}\sigma^{\mu\nu}N ]
              \ ,            \label{eq:L1again}
\end{eqnarray}
where the constant
\begin{equation}
 C \equiv{2\over 9}  {h_{\rm A}^2\over M_\Delta^2}
           [ (Z+\case1/2)^2(2M_\Delta+M)-(Z+\case1/2)M_\Delta ]
                             \ . \label{eq:Y}
\end{equation}
The second and third terms on the right of (\ref{eq:L1again}) 
can clearly be absorbed into the corresponding 
terms in ${\cal L}_{\pi N}$ of (\ref{eq:LpiN}). For example,
we can define $\beta_\pi = \beta'_\pi - M C$
so that only $\beta'_\pi $ is physically significant.
Thus, one arrives at  a
nonlocal Lagrangian for
the $\pi N$ system  given by
\begin{equation}
  {\cal L}_{\equiv}(x) =
                 {\cal L}_{\pi N}(x)
      - h_{\rm A}^2  \overline{N}
             \bbox{T\cdot a}_\mu S^{\mu\nu}
            \bbox{T}^\dagger\bbox{\cdot a}_\nu N 
               \ . \label{eq:effagain}             
\end{equation}
Since this $Z$-independent Lagrangian is equivalent
to that in (\ref{eq:L}), we conclude that the
$Z$ parameter is physically irrelevant.

We note, however, that it is not convenient to 
work with a nonlocal Lagrangian as in (\ref{eq:effagain}) since
it is hard to maintain the various symmetries
and to organize the Lagrangian according to the naive dimensional analysis
\cite{GEORGI}. Although one may still expand the second term 
in (\ref{eq:effagain})
in a Taylor series in derivatives of the pion field, each
derivative will be suppressed by $1/(M_\Delta-M)$ rather than $1/M$.
This cannot be a useful expansion since it violates  the naive
dimensional analysis\cite{GEORGI} and any truncation would result
in large errors.
This shows that
one needs to treat the $\Delta$ as a dynamical degree of freedom
in agreement with Refs.~\cite{BAN_MIL95,JENKINS91b,BIRA}.
Thus our analysis indicates that the original effective Lagrangian
(\ref{eq:L}) should be used including the $\Delta$ field explicitly and
making some convenient choice 
for the parameters $A$ and $Z$ (such as $A=-1$, $Z=-1/2$) since they are not 
relevant to the physics.

For clarity we have limited our discussions to terms with two derivatives
on the pion field, our results can be generalized nevertheless.
Consider the $\pi N\Delta$ or $\pi\Delta\Delta$ vertex
in (\ref{eq:lagDel}). If it is 
attached to a $\Delta$ line in a Feynman diagram of an arbitrary order,
the pole of the $\Delta$ propagator will be removed
according to (\ref{eq:red}) if 
the vertex contains a $\gamma_\mu$ that is contracted with the propagator.
This is the case for the $Z$, $Z_2$, and
$Z_3$ terms where the $\Delta$ line will shrink to a point
and  generate local vertices which can be combined with similar
vertices in the Lagrangian. Thus the $Z$, $Z_2$, 
and $Z_3$ parameters can all be subsumed in the 
infinite number of parameters in the Lagrangian to all orders.

To summarize, we have demonstrated that the $\Delta$ should be 
treated as a dynamical degree of freedom 
and all the off-shell parameters $Z$, $Z_2$, and $Z_3$ are redundant
since they can be absorbed into other parameters in the Lagrangian.
Our results should pave the way for systematic inclusion of $1/M$
corrections involving the $\Delta$ isobar in heavy-baryon ChPT and
in calculations for nuclear matter and finite nuclei where
the $1/M$ corrections are crucial.

We thank D.K. Griegel, S. Jeon, J.I. Kapusta and
B.D. Serot for useful comments and 
stimulating discussions. We acknowledge support
from the Department of Energy under grant No. DE-FG02-87ER40328.

\appendix
\section*{}

To be self-contained, we briefly discuss the transformation of the
various fields under a non-linear realization\cite{CCWZ} 
of $SU(2)_{\rm L}\otimes SU(2)_{\rm R}$ 
symmetry. More details may be found in
e.g. Ref.\cite{FSTprepare}.
With $L \in SU(2)_{\rm L}$ and $R \in SU(2)_{\rm R}$, one has
the mapping
\begin{equation}
L\otimes R:\ \ \ (\xi, N, \Delta_\mu)\longrightarrow 
        (\xi', N', \Delta'_\mu)   
          \ ,     \label{eq:nonlr}
\end{equation}
where
\begin{eqnarray}
\xi'(x) &=& L \xi(x) h^{\dagger}(x) = h(x) \xi(x) R^{\dagger}
               \ , \label{eq:Xitrans} \\[4pt]
 N'(x) &=& h(x)N(x)  \ , \label{eq:Ntrans} \\[4pt]
 \bbox{\Delta}'_\mu(x) &=& \thalf 
              h \, {\rm tr}(h\bbox{\tau}
            h^\dagger \bbox{\tau})\bbox{\cdot} \bbox{\Delta}_\mu (x)
              \ .       \label{eq:Dtrans}
\end{eqnarray}
The second equality in (\ref{eq:Xitrans}) defines
$h(x)$ as a function of $L$, $R$, and the local pion field:
$
h(x)=h(\bbox{\pi}(x),L,R) \ .
$
The pseudoscalar nature of the pion field 
implies then $h(x)\in SU(2)_{\rm V}$, with
$ SU(2)_{\rm V}$ the unbroken vector  subgroup
of $SU(2)_{\rm L} \otimes SU(2)_{\rm R}$.
Eqs.~(\ref{eq:Ntrans}) and (\ref{eq:Dtrans})  
ensure that the nucleon  transforms linearly under
$SU(2)_{\rm V}$ as an isospin Pauli-spinor and  the $\Delta$ 
as an isospin vector Pauli-spinor. (Note that 
$\thalf {\rm tr}(h\bbox{\tau}h^\dagger \bbox{\tau})$ provides the 
orthogonal transformation of the isovector part of the field).
Writing out the $\bbox{\Delta}$ field explicitly, one finds
\begin{eqnarray}
     \Delta_\mu^1 + i  \Delta_\mu^2 & = &
             {1\over \sqrt{3}} \left( \!\!
                    \begin{array}{c}
                      \sqrt{2} \Delta_\mu^{0}\\
                      \sqrt{6} \Delta_\mu^{-}
                    \end{array}  \!\!
             \right)
                     \\[3pt]
    \Delta_\mu^1 - i  \Delta_\mu^2 & = &
            - {1\over \sqrt{3}} \left( \!\!
                    \begin{array}{c}
                      \sqrt{6} \Delta_\mu^{++}\\
                      \sqrt{2} \Delta_\mu^{+}
                    \end{array}  \!\!
             \right)
                     \\[3pt]
     \Delta_\mu^3 &=& \sqrt{2\over 3} \left( \!\!
                    \begin{array}{c}
                      \Delta_\mu^{+}\\
                      \Delta_\mu^{0}
                    \end{array}\!\!
             \right)       \ .
\end{eqnarray}

It is useful to note the following properties of the isospin
transition matrix defined in (\ref{eq:T}):
\begin{equation}
     T_a^\dagger T_a = \bbox{1}
           \ , \ \ \ \ \ \
      T_a T^\dagger_b = \delta_{a b} - {1\over 3} \tau_a \tau_b
              \ ,
\end{equation}
where $\bbox{1}$ represents the $4\times 4$ unit matrix.
Also the
transformation properties of the axial vector and vector fields
are
$a_{\mu}'=h a_{\mu}h^{\dagger}$ and $v_{\mu}'=hv_{\mu}h^{\dagger}
-ih\partial_{\mu}h^{\dagger}$.

\end{document}